\documentstyle[prl,aps,preprint]{revtex}
\tightenlines
\begin{document}
\draft
\title{Phonon assisted tunneling in Josephson junctions}
\author{E.G. Maksimov, P.I. Arseyev and N.S. Maslova}
\address{P.N. Lebedev Physical Institute RAS, Moscow 11\,7924, Russia}
\date{\today}
\maketitle
\begin{abstract}
The expression for additional subgap current in the presence of
electron--phonon
interaction is derived. We show that the phonon assisted tunneling leads to the
appearance of peaks on current-voltage characteristics at the
Josephson frequencies corresponding to the
Raman-active phonons. The relation of the obtained results to the experimental
observation are discussed.
\end{abstract}


It is well known that the superconducting Josephson current at nonzero
voltage~$V$ across the junction oscillates at a so called Josephson
frequency~$\omega$ \cite{ref1}:
\begin{equation}
   \hbar\omega = 2eV
   \label{neq1}
\end{equation}
were $V$ is the voltage on the contact.

This alternating (AC) supercurrent can generate time dependent
longitudinal or transverse electromagnetic
fields within the junction. Nonlinear character of this phenomenon leads to the
appearance of many  peculiarities in the current--voltage characteristic (CVC)
of a Josephson contact \cite{ref2}. As an example, the CVC can exhibit current
peaks, or multiple-valued CVC with different branches can appear at some
voltages. These effects
result from the coupling of the Josephson oscillation to electromagnetic
cavity modes
of the junction structure \cite{ref3,ref4}.
Voltages are connected with the corresponding mode frequencies $\omega_i$
by the Josephson relation Eq.\ (\ref{neq1}).
The possibility of the generation
of phonons in Josephson junctions have been predicted by Ivanchenko and
Medvedev \cite{ref5}. The authors have considered the excitation of the long
wavelength acoustic resonance modes in the dielectric layer of the contact,
which can influence the shape of the CVC of the junction in the same way
as excitation of electromagnetic cavity modes.
To our knowledge,
the resonant current peaks in the CVC and the branching of the CVC,
corresponding to the acoustic resonances predicted in \cite{ref5}, have never
been observed experimentally. The mere fact of the phonon generation in
classical Josephson junctions, caused by the quasiparticle current or by the
AC Josephson current, was repeatedly found experimentally
\cite{ref6,ref7,ref8}. This generation was observed by measuring phonon
emission using phonon spectrometer.
Two mechanisms of this generation were discussed:
the direct conversion of electromagnetic waves into phonons
and phonon-assisted tunneling processes \cite{ref8}.
Which of the mechanisms better fits
experimental situation  have not been established at full length.

Recently specific subgap structures in the form of current peaks (resonances)
have been observed \cite{ref9,ref10} in the CVC\rq s of intrinsic Josephson
junctions in $\mathrm{Bi_2Sr_2Ca_1Cu_2O_8}$ and
$\mathrm{Tl_2Ba_2Ca_2Cu_3O_{10}}$ with the tunneling current in the
$\vec{c}$-direction, perpendicular to the copper--oxygen layers.
The obtained results have been explained\cite{ref11,ref11a} by the coupling of
the optical phonons in the dielectric layers being the ionic crystal
with the
AC Josephson current
(or, more precisely, with the longitudinal electric field existing
inside the dielectric layer).
The displacement current contribution ($\dot D $)
to the total current
was described in the framework of a simple macroscopic dielectric function for
ionic crystals $\varepsilon(\omega)$.
It was shown that peaks in the CVC appear when the voltage
satisfies the condition
\begin{equation}
   2eV = \hbar\omega_{\rm LO}
   \label{bneq2}
\end{equation}
where $\omega_{\rm LO}$ is the frequency of the longitudinal optical (LO)
phonons
with the wavevector $q=0$.
These results can be easily obtained from the equations of Refs.
\onlinecite{ref3,ref4} for the electromagnetic waves generation
if one substitutes the displacement current $\dot D $ in the form:
\begin{equation}
   \dot D(t)= i\! \int\!\! {\rm d}\omega \omega \varepsilon(\omega)E(\omega)
  {\rm e}^{i\omega t}
\end{equation}
The peak appearance corresponds to
zero value of the dielectric function, because $\varepsilon(\omega_{LO})=0$.

A fine structure in the CVC of a single $\mathrm{Bi_2Sr_2Ca_1Cu_2O_8}$ break
junction have been reported earlier \cite{ref12,ref13} which could have some
resemblance with the subgap structure in BSCCO stacks \cite{ref9,ref10,ref11}.
The results of the detailed investigations of this structure are presented in
the preceding article \cite{ref14}. The essential result of this work is the
observation of the local branching of the CVC and the dips in the
${\rm d}I/{\rm d}V$ characteristics at the voltages~$V$ satisfying the condition
\begin{equation}
   2eV = \hbar\omega_{\rm R}
   \label{neq5}
\end{equation}
where $\hbar\omega_{\rm R}$ are the energies of the Raman-active phonons.

The explanation for the observed subgap structure given in
Refs.\onlinecite
{ref11,ref11a} is based on  direct conversion of the AC longitudinal
electric field, existing {\em inside} the junction into LO phonons carrying
electric field also {\em inside} the junction.
But the direct conversion of longitudinal electric field
into Raman-active phonons  is forbidden due to selection rules.
Moreover, the very existence of the well defined dielectric layers
in the breakjunctions studied in Ref.\onlinecite{ref14} is very
improbable.
Thus "electromagnetic" explanation can not be applied to the case of
 Raman active phonons, but such phonons  can be emitted due to the
electron-phonon interaction
in the banks of the junction
and can interfere with AC Josephson current.

Here we apply the ideas developed in the work \cite{ref5} to explain the
results obtained in Refs. \onlinecite{ref12,ref13,ref14}.
For this purpose we generalize the
approach used in \cite{ref5} to include interaction of electrons with all phonons
and not only with acoustical ones.
This generalization is very important for the high-Tc materials, because
the range of Josephson frequencies exceeds the range of the main part of
the phonon spectra.
The total Hamiltonian can be written for
this case as
\begin{equation}
   {\mathcal{H}}_{\rm{total}} = {\mathcal{H}}_1 + {\mathcal{H}}_2
         + \left( \sum_{k, p, \sigma} T_{kp}\,a^{+}_{k\sigma}\,b_{p\sigma}
         + {\rm h.c.} \right)
         + \left( \sum_{k,p,\sigma \atop q,\lambda}
           T_{kp}\,a^{+}_{k\sigma}\,b_{p\sigma}\,\alpha_{q\lambda}\,
           (c_{q\lambda} + c^{+}_{-q\lambda}).
         + {\rm h.c.} \right)
	   \label{neq6}
\end{equation}
Here the terms ${\mathcal{H}}_1$ and ${\mathcal{H}}_2$ describe electrons on
the two sides of the tunneling junction with the operators $a_{k\sigma}$ and
$b_{p\sigma}$ correspondingly. These Hamiltonians include
an interelectron interaction leading to superconductivity. We do not
consider here this interaction explicitly. It can be the electron--phonon
interaction or any other.
The only important assumption made is that this
interaction leads to the appearance of superconductivity and
the Josephson current exists between the two superconductors.
We will use the BCS formulas for Josephson current for the case of
isotropic order parameter in the main part of the paper.
Qualitatively, results are not very sensitive to the symmetry of
pairing.
It is supposed also that these Hamiltonians
include the interaction of the electrons with the bias voltage
\begin{equation}
   -eV_1(t)\sum_{k,\sigma}a^{+}_{k\sigma}a_{k\sigma}
   \quad \mbox{and} \quad
   -eV_2(t)\sum_{p,\sigma}b^{+}_{p\sigma}b_{p\sigma}
   \label{neq7}
\end{equation}
where
\begin{equation}
   V_1(t) - V_2(t) = V(t) \label{neq8}
\end{equation}
and $V(t)$ is the bias voltage, applied to the junction.
Since we are interested in direct current changes, we can use the usual
approximation  simplifying the problem. Namely, we neglect
electrodynamic selfinteraction in the junction, which is quite reasonable
for weak AC component of the current (Josephson current) \cite{ref2}.
Then the bias voltage  is equal to a given constant external bias $V(t)=V$.
The third term in
Eq.\,(\ref{neq6}) is the usual tunneling Hamiltonian and $T_{kp}$ is a matrix
element for the tunneling of an electron. The fourth term in
Eq.\,(\ref{neq6}) describes the phonon assisted tunneling of electrons.
The value $\alpha_{q\lambda}$ stands for the
electron--phonon coupling and the operators $c_{q\lambda}$ and
$c^{+}_{-q\lambda}$ describe phonons with the given momentum $q$ and the
polarization $\lambda$. A similar expression for this term was used earlier by
Kleinman \cite{ref16} to calculate the contribution of the phonon assisted
tunneling to the quasiparticle current between superconductors. We would like to
emphasize here that the existence of such a term in the total Hamiltonian does
not depend on the origin of the superconductivity in HTSC systems. There is some
experimental evidence \cite{ref17} that the HTSC material possesses
considerably strong electron--phonon interaction even though it does not govern
totally the superconducting properties of HTSC systems.

Introducing the expression for the total current as
\begin{equation}
   I = e\langle\dot{N}_1\rangle
   \label{neq10}
\end{equation}
where $N_1$ is the number of electrons in one side of the junction and $e$ is
the electron charge
\begin{equation}
   N_1= \sum_{k,\sigma} a^{+}_{k\sigma}a_{k\sigma}.
   \label{neq11}
\end{equation}
We can write for the current the following expression ($\hbar =1$)
\begin{equation}
   I = -{\rm i}e\langle[N_1, \mathcal{H}]\rangle.
   \label{neq12}
\end{equation}
The symbol $\langle\cdots\rangle$ means the averaging over the nonequilibrium
statistical ensemble.
The last two terms in
Eq.\,(\ref{neq6}) give the nonzero contribution to this commutator.
 Substituting these terms into Eq.\,(\ref{neq12}) we get
\begin{equation}
   I = 2e\,{\rm Im}\left\{\sum_{k,p,\sigma} T_{pk} \left\langle
       a^{+}_{k\sigma}(t) b_{p\sigma}(t) \right\rangle
       + \sum_{k,p,\sigma \atop q,\lambda}
           T_{pk}\alpha_{q\lambda} \left\langle
           a^{+}_{k\sigma}(t) b_{p\sigma}(t)
           \left(c_{q\lambda}(t) + c_{-q\lambda}^+(t)
           \right)\right\rangle\right\} .
       \label{neq13}
\end{equation}
After long but simple calculations following close to that done in the work
\cite{ref5} we get
\begin{equation}
   I(t) = \left( 1 + \sum_{q,\lambda} \alpha_{q\lambda}
           \left\langle c_{q\lambda}(t) + c_{-q\lambda}^+(t) \right\rangle
          \right)
           \left[ I_S(t) + I_N \right]
           \label{neq14}
\end{equation}
where $I_S(t)$ describes the contribution of the superconducting Josephson
current (AC part) to the total current and $I_N$ (DC part)
gives the contribution of the
quasiparticle current. The Josephson current can be written in the form
\begin{equation}
   I_S(t) = {\rm Re}\left\{ I_p\left(\frac{\omega_j}{2} \right)\right\}
                    \sin\phi(t) +
            {\rm Im}\left\{ I_p\left(\frac{\omega_j}{2} \right)\right\}
                    \cos\phi(t)
            \label{neq15}
\end{equation}
where
\begin{equation}
   \omega_j = \frac{2eV}{\hbar}.
   \label{neq17}
\end{equation}
The value $\phi(t)$ is the order parameter
phase difference between the two sides of the junction:
\begin{equation}
   \phi(t) = \omega_jt.
   \label{neq18}
\end{equation}
We shall not reproduce here the expressions for the values
$I_p\left(\frac{\omega_j}{2}\right)$ and
$I_N$ which can be found in the works
\cite{ref2,ref4,ref17a}. It is only important for us that
${\rm Im}\left\{I_p(\omega_j)\right\}$ and
$I_N(V)$ are very small at low temperature~$T$
and at small voltage~$V$. For the BCS model
${\rm Re}\left\{I_p(\omega)\right\}$ at small voltages has the form
\begin{equation}
   {\rm Re} \left\{I_p \left(\frac{\omega_j}{2} \right)\right\}
   \approx I(0) \approx
   \frac{\pi}{2} \frac{\Delta (T)}{e\, R_{\rm N}} \tanh
   \frac{\Delta (T)}{2\,T}.
   \label{neq19}
\end{equation}
Here $\Delta$ is the superconducting gap, $e$ is the electron charge and
$R_{\rm N}$ is the resistivity of the contact in the normal state. As
$V$ increases the current amplitude
${\rm Re} \{I_p(\omega)\}$ increases also and
at $V =2\frac{\hbar}{e} \Delta$ it
has the Riedel logarithmic singularity \cite{R} (qualitatively originated
from the density of states singularity at the gap edges).

To calculate the total current in the presence of the phonon
assisted tunneling we should calculate the value
$\langle c_{q\lambda} + c^{+}_{-q\lambda} \rangle$. Using the
Hamiltonian (\ref{neq6}) and adding the free phonon part $\mathcal{H}_{\rm ph}$
\begin{equation}
   {\mathcal{H}}_{\rm ph} = \sum_{q,\lambda} \omega_{q\lambda}
                            c^{+}_{q\lambda} c_{q\lambda}
                            \label{neq20}
\end{equation}
we can obtain the equation of motion in Heisenberg representation
for the phonon operator
$(c_{q\lambda}(t) + c^{+}_{-q\lambda}(t))$. For the averaged values
this gives:
\begin{equation}
   {\mathcal{D}}^{-1}_{q\lambda}(t)
   \left\langle c_{q\lambda}(t) + c^{+}_{-q\lambda}(t) \right\rangle
   =
   -2{\rm Re} \sum_{k,p,\sigma}
   \int\!\!{\rm d}r\,\,{\rm e}^{-{\rm i}qr}\,\alpha_{q\lambda} T_{kp}
   \left\langle (a^{+}_{k\sigma}(t) b_{p\sigma}(t) \right\rangle.
   \label{neq21}
\end{equation}
Here ${\mathcal{D}}^{-1}_{q\lambda}(t)$ is operator inverse to
the phonon Green function
\begin{equation}
   {\mathcal{D}}^{-1}_{q\lambda}(t) =
  - \frac{1}{\omega_{q\lambda}} \left[ \frac{\partial^2}
    {\partial t^2} +  \omega ^2_{q\lambda} + \gamma \frac{\partial}
    {\partial t} \right]
   \label{neq22}
\end{equation}
where $\omega_{q\lambda}$ are the phonon frequencies and $\gamma$ is the
small phonon relaxation rate.
The right-hand side of Eq.\ (\ref{neq21}) appears due to phonon-assisted
tunneling processes and plays the role of an external source for
lattice vibrations. (For the equilibrium case
$\langle c_{q\lambda}(t) + c^{+}_{-q\lambda}(t) \rangle =0 $ ).
Following again the work
\cite{ref5} we can calculate the contribution to this function from
oscillating Josephson current:
\begin{equation}
   \sum_{k,p}
   \int\!\!{\rm d}r\,\,{\rm e}^{{\rm i}qr}\,\alpha_{q\lambda} T_{kp}
   \left\langle a^{+}_{k\sigma}(t) b_{p\sigma}(t) \right\rangle
   =
   \frac{\alpha_{0\lambda}}{e}\,
   \left(I_p\left(\frac{\omega_j}{2}\right)
   {\rm e}^{{\rm i}\phi(t)}\right) \delta_{q,0}
     \label{neq23}
\end{equation}
Since we have neglected the space dependence of the phase
$\phi (t)$ only the $q=0$ component gives a nonzero contribution in
Eq.\,(\ref{neq23}).

The solution of Eq.\,(\ref{neq21}) can be easily found now:
\begin{equation}
   \left\langle c_{0\lambda}(t) + c^{+}_{0\lambda}(t) \right\rangle
   =
   -\frac{2\alpha_{0\lambda}}{e}\,
   {\rm Re}\left\{{\mathcal{D}}_{0\lambda}\left(\omega_j\right)
   I_p\left(\frac{\omega_j}{2}\right)
   {\rm e}^{{\rm i}\omega_jt}\right\}
   \label{neq24}
\end{equation}
with phonon Green function ${\mathcal{D}}_{0\lambda}\left(\omega \right)$:
\begin{equation}
   {\mathcal{D}}_{0\lambda}(\omega) =
   \frac{\omega _{0\lambda}}{\omega ^2 - \omega ^2_{0\lambda}
   - {\rm i}\omega\gamma}
   \label{D}
\end{equation}
Now we consider for  simplicity the case of low temperature~$T$, so
only the terms with Re $\{ I_p \} $ should be taken into account.
In this case Eq.\,(\ref{neq24}) can be written as
\begin{equation}
   \left\langle c_{0\lambda}(t) + c^{+}_{0\lambda}(t) \right\rangle
   =
   2\frac{\alpha_{0\lambda}\omega_{0\lambda}}{e}\,
     {\rm Re} I_p\left(\frac{\omega_j}{2}\right)
\Bigl[
 {\rm Im}\left\{{\mathcal{D}}_{0\lambda}\left(\omega_j\right)\right\}
       \sin\omega_jt
- {\rm Re}\left\{{\mathcal{D}}_{0\lambda}\left(\omega_j\right)\right\}
        \cos\omega_jt
\Bigr]
   \label{neq25}
\end{equation}
Substituting Eq.\,(\ref{neq25}) into Eq.\,(\ref{neq14}) and averaging
the oscillating terms
over the time, we
get the expression for the  excess current
$\Delta I(V)$. Using the explicit form of
${\mathcal{D}}_{0\lambda}\left(\omega_j\right) $  (Eq.\ (\ref{D}))
we finally obtain :
\begin{equation}
   \Delta I(V) = \sum_{\lambda}\,
                 \frac{\alpha^2_{0\lambda}\omega_{0\lambda}}{e}\,
                 \frac{\gamma\omega_j \left\{{\rm Re}I_p(\omega_j/2)\right\}^2}
                      {\left(\omega^2_j - \omega^2_{0\lambda} \right)^2
		       + \gamma^2\omega^2_j}
   \label{neq26}
\end{equation}
where $\omega_j$ is defined by Eq.\,(\ref{neq17}).
This expression shows that there are peaks in the CVC at the voltages
corresponding to the condition
\begin{equation}
   2eV = \hbar\omega_{0\lambda}
   \label{neq27}
\end{equation}
where $\omega_{0\lambda}$ are the frequencies of the optical phonons with $q=0$.
The value ${\rm d}I(V)/{\rm d}V$ has correspondingly the dips very close to
voltages satisfying Eq.\,(\ref{neq27}). The difference between the peaks
positions and the dips is of the order of a small value~$\gamma$. Note,
that contrary to Eq. (\ref{neq1}), now Eq. (\ref{neq27}) includes all
optical phonons (not only LO) and these are the phonons in superconductors
themselves and not in the intermediate dielectric layer. The value
of the electron--phonon coupling~$\alpha^2_{0\lambda}$
determines how great the contribution from each of the phonon
branches is.
This value is equal to zero for
some phonon modes with $q=0$ due to the symmetry reason. This is the case,
for
example, for infrared-active modes. As it is well known \cite{ref18} the
coupling constant of electrons with the Raman-active phonons is nonequal to zero
even for $q=0$. Moreover, as it was shown by the first principle calculations
\cite{ref19} the constant of coupling with Raman-active phonons is considerably
large. This is why the peaks mainly connected with such phonons are observed in
the experiments made on break junctions \cite{ref14}.

The current excess can be find for all temperatures~$T$ and voltages~$V$
numerically using Eq.\,(\ref{neq14}) and Eq.\,(\ref{neq24}) obtained in
this work. In the present work we would like only to mention some clear
qualitative
consequences of such  calculations. The peak amplitudes --- or the dips in
the  ${\rm d}I(V)/{\rm d}V$ characteristics --- will demonstrate the Riedel
peculiarities: the amplitude will increase for the values of $eV/\hbar$
close to $2\Delta$. This subgap structure will disappear for
$eV/\hbar$ larger than $2\Delta$, because the value $I_p(V)$
quickly goes to zero for $eV/\hbar > 2\Delta$.
The amplitude of the peaks decreases
with decreasing of the critical Josephson current as $I^2_c(0)$.
We would like to emphasize here that the amplitude of the peaks predicted in the
work \cite{ref11,ref11a} is also proportional to $I^2_c$ in contradiction with
the statement announced in these works. These amplitudes will also change in the
external magnetic field due to the change of the value~$I_c$.

In conclusion,
simple microscopic theory
describing the phonon assisted Josephson
tunneling was derived. It was shown that there exists subgap
excess current.
The current excess demonstrates peaks at voltages $V$ such as
$2eV/\hbar$ is equal to the frequencies of the Raman-active phonon modes. The
amplitudes of the peaks are proportional to $I^2_c$, the square of the Josephson
critical current. All these results are in good agreement with the experimental
work \cite{ref14}.

The case of anisotropic pairing (e.g. $d$--pairing) will not
change essentially the obtained results, till the AC Josephson current
exists.
The functions $I_p(\omega)$ and $I_S(\omega)$ will
be certainly different from
that of given in the Ref.\,\cite{ref2,ref4,ref17a}, but all other equations will
not be changed.
The only necessary things are the existence of the AC Josephson current
and the electron-phonon interaction. Therefore the position of the subgap
structures on the CVC is determined in any case by
the frequencies of Raman-active
modes. Of course, the form of this structure and
its temperature dependence can differ
from those for isotropic pairing. The problem of numerical calculations of
CVC for $T \ne 0$ in different pairing models is now under investigation.

\acknowledgments
   This work was supported partially by RFBR grants 96--02--16134 ,
   96--15--96476 and grant 96-081 of the Programm "Superconductivity".
   One from the authors (E.\,G.\,M.) would like to thank the
   University of Wuppertal and Prof.\,H.\,Piel for kind hospitality.
   We are grateful to Prof.\,Ya.\,Ponomarev for the communicating the
   results \cite{ref14} to us before publication and for the
   continuous exchange of information.


%
%

%
%

\end{document}